# Acoustic diagnostics of femtosecond laser filamentation

**BINPENG SHANG,** [1,2] **NAN ZHANG,** [1,2] **PENGFEI QI,** [1,3] **SHISHI TAO,** [1,2] **LIE LIN,** [1,3] **AND WEIWEI LIU** [1,2]

[1] *Institute of Modern Optics, Nankai University, Tianjin 300350, China*
[2] *Tianjin Key Laboratory of Micro-scale Optical Information Science and Technology, Tianjin 300350, China*
[3] *Tianjin Key Laboratory of Optoelectronic Sensor and Sensing Network Technology, Tianjin 300350, China*

**Abstract:** The promising application of femtosecond laser filamentation in atmospheric remote sensing brings imperative demand for diagnosing the spatiotemporal dynamics of filamentation. Acoustic emission (AE) during filamentation opens a door to give the insight into the dynamic evolution of filaments in air. In particular, the frequency features of the acoustic emission provide relevant information on the conversion of laser energy to acoustic energy. Here, the acoustic emission of femtosecond laser filament manipulated by energy and the focal lengths was measured quantitatively by a broadband microphone, and the acoustic parameters were compared and analyzed. Our results showed that the acoustic power presents a squared dependence on the laser energy and the bandwidth of the acoustic spectrum showed a significant positive correlation with laser energy deposition. It was found that the spectrum of the acoustic pulse emitted from the middle of the filament has a larger bandwidth compared to those emitted from the ends of the filament and the spectrum of the acoustic pulse is also an indicator of the filament intensity distribution. These findings are helpful for studying the plasma filament properties and complex dynamic processes through acoustic parameters and allow the optimization of remote applications.

## 1. Introduction

Filamentation is formed by the dynamic balance between Kerr self-focusing and plasma defocusing during femtosecond laser pulse propagation in transparent dielectrics [1, 2]. A plasma filament can sustain hundreds of meters and up to several kilometers in the ambient atmosphere without an external focusing element [3, 4]. The long-distance propagation capability and robustness of filament are expected to realize many potential applications, such as remote sensing and atmospheric pollution detection [5-7]. Powerful filaments with an intensity of about $10^{13}$ W/cm$^2$ can effectively ionize, break, and excite the molecules and fragments, resulting in specific fingerprint fluorescence spectra, which provide a great opportunity for real-time multi-component detection. To satisfy different requirements in these practical applications, manipulating long-distance filamentation of femtosecond pulses has attracted intensive interest, and brings imperative demand for diagnosing filamentation characteristics [8-11].

It is a challenge to diagnose the high-intensity zone produced by femtosecond laser pulse filamentation [12], since almost any material inserted into a high-intensity laser beam will be destroyed. By a feat of the induced abundant physical effects during filamentation, various indirect methods such as optical interferometry [13], acoustic [14, 15], imaging [16], fluorescence [17], and plasma conductivity [18] have been utilized to diagnose the plasma. Due to the advantages of fast, non-destructive, and high-resolution, the acoustic diagnosis based on observable photoacoustic effects demonstrates tremendous potential in measuring filamentation.

Air ionization by focusing femtosecond laser pulses of sufficient energy leads to the generation of strong shock waves that progressively transform into acoustic waves as they propagate outwards from the focusing area [19, 20]. The properties of laser-generated acoustic waves, such as duration, energy, frequency spectrum, etc., depend on the laser pulse parameters (pulse duration, energy, wavelength), and focusing conditions. Generally, laser-induced acoustic signals represent the total plasma inhomogeneity due to laser-matter interactions [21] and are used to estimate the electron densities of plasma [22]. These acoustic waves emitted from the plasma contain information about plasma expansion [23], making this method useful for filament-based remote sensing. These acoustic waves studies were used to understand and monitor the laser-matter interaction [24], especially in the energy conversion from laser to acoustic spectrum via plasma expansion. However, existing studies generally focus on the spatial distribution of filament intensity as characterized by acoustic amplitude and the analysis of plasma formation mechanisms on different metal surfaces by acoustic spectrum [14, 25, 26]. There is still a lack of systematic quantitative analysis of the dynamic evolution and energy conversion process of filaments.

In this paper, we present our work on the acoustic diagnostics of the femtosecond filament generated using four different focusing conditions ($f$ =5 cm, 5 m, 10 m, and 20 m). The acoustic emission of femtosecond filament manipulated by energy and the focal lengths were measured quantitatively by a broadband microphone. The results showed that acoustic power presents a squared dependence on the laser energy. Acoustic spectral features related to the laser energy deposition were identified and discussed. The spectrum of the acoustic pulse emitted from the middle of the filament has a larger bandwidth compared to those emitted from the ends of the filament. For a given laser energy, the acoustic pulse with the maximum bandwidth appears at the focal condition with medium focal length. These findings provide a guiding significance for understanding and monitoring the plasma filament dynamic processes and lay a firm foundation for femtosecond laser filament-based remote applications.

## 2. Experimental setup

The experimental setup is shown in Fig. 1. A commercial Ti: sapphire femtosecond laser amplifier (Legend, Coherent Inc.) was used to deliver 50 fs, 800 nm pulses with an energy up to 5 mJ at a repetition rate of 500 Hz. The combination of a half-wave plate (HWP) and a polarizing beam splitter (PBS) attenuates the laser energy to the desired value. For generating filaments within the allowable range in the laboratory (up to 30 m), the initial Gaussian laser beam (8 mm diameter at the $1/e^2$ level) was focused by the combination of a plano-concave lens L1 ($f=-75$ mm) and a plano-convex lens L2 ($f$ =500 mm).

The acoustic emission (AE) from the filament is recorded by a microphone (V306, Olympus. Ltd.) with a bandwidth of 4.5 MHz. The distance between the filament and microphone is set to 10 mm to avoid signal saturation. During the measurement, the microphone was mounted on an electronically controlled translation stage and moved parallel to the filament. An ultrasonic pulse receiver (5072PR, Olympus. Ltd.) was used to amplify the acoustic signal detected by the microphone and the amplified signal was acquired by an oscilloscope (DPO3034, Tektronix Inc.). Each acoustic signal presented in this paper is the result of an average of 256 measurements in order to improve the signal-to-noise ratio.

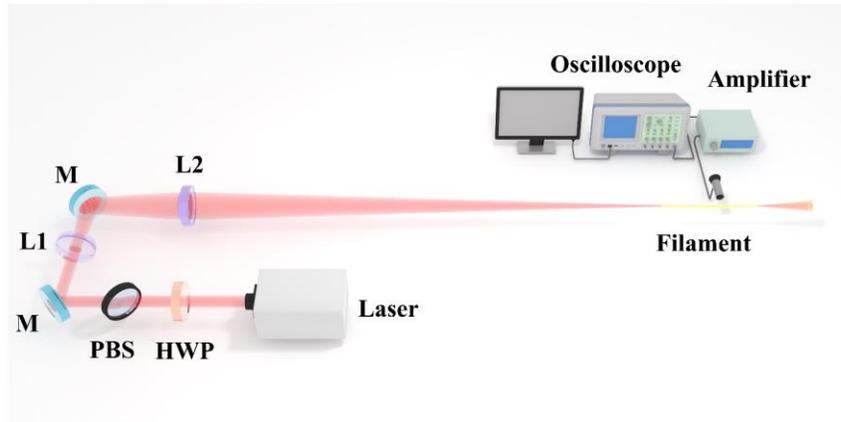

Fig. 1. Experimental setup for the detection of acoustic emission from femtosecond laser filament.

## 3. Results and discussion

The pulse energy dependence of the temporal acoustic signal generated by the femtosecond laser filament was investigated for different focal lengths ranging from 5 cm to 20 m. The acoustic signal is measured in the middle of the filament and the microphone has a spatial resolution of 0.87 cm [27]. The experimental results are shown in Fig. 2. As can be observed in Fig. 2a, the acoustic pulse duration is roughly 15 µs at a focal length of 5 cm. The waveform consists of 2~3 main oscillatory cycles followed by a series of small ripples. The acoustic pulse duration and the number of main oscillation cycles decrease as the focal length increases. The peak-to-peak value of the acoustic signal is the oscilloscope voltage value (mV). As shown in Fig. 2b-2e, the acoustic time domain waveform does not change significantly with increasing pulse energy for a certain focal length. Usually, an increase in the laser energy leads to a rise in the acoustic amplitude [28], which is due to the increase of the laser energy deposited in the filament.

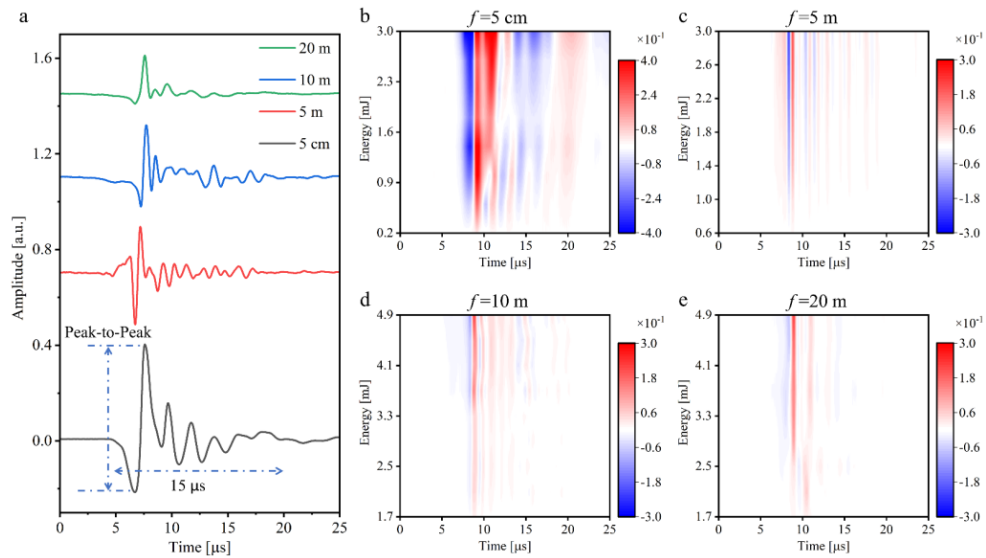

Fig. 2. (a) Acoustic time domain waveforms measured by microphone and dependences of the temporal profile of the acoustic wave on the laser pulse energy for different focal lengths of 5 cm (b), 5 m (c), 10 m (d), and 20 m (e).

Fig. 3 shows the pulse energy dependence of the acoustic amplitude for different focal lengths. It is seen in all plots that the acoustic amplitude increases linearly with the pulse energy for all the focal lengths $R^2 \geq 0.964$. The acoustic power can be calculated by $\overline{W} = \beta^2 f \sum [V(t)]^2 \Delta t / (\rho_0 c_0)$, where $\beta = P(t)/V(t)$ is the responsivity of the microphone and nearly a constant in the bandwidth from 50 Hz to 100 kHz, $V(t)$ is the voltage recorded by the oscilloscope, $f$ is the repetition rate of the femtosecond laser, $\Delta t$ is the sampling interval of the oscilloscope, $\rho_0$ and $c_0$ are air density and acoustic velocity, respectively. Obviously, acoustic power follows a squared dependence with laser energy (or intensity I) $\overline{W} \propto I^2$. This relation is consistent with the gas absorption via non-resonant two-photon rotational Raman excitation in nitrogen and oxygen [29, 30]. During the filamentation, a femtosecond laser pulse propagates through the atmosphere, it mainly deposits energy into the propagation medium through optical field ionization and non-resonant rotational Raman excitation of the air molecules [31]. Therefore, it infers that the energy deposited by the optical field ionization is dissipated mainly through the fluorescent radiation, while the deposited energy via the latter mechanism mainly dissipates by emitting the acoustic pulse.

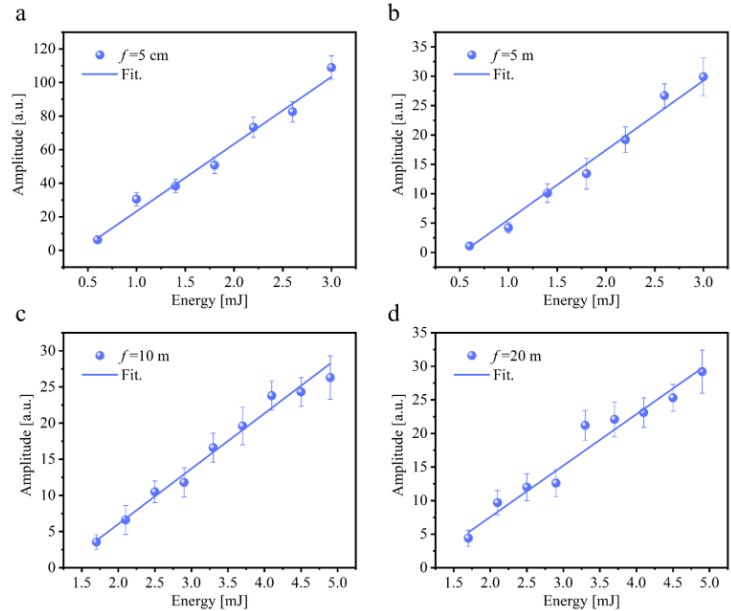

Fig. 3. Dependences of the acoustic amplitude on the laser pulse energy for different focal lengths of 5 cm (a), 5 m (b), 10 m (c), and 20 m (d).

The dependence of the acoustic pulse spectrum on the focal length is also investigated. The acoustic spectral features provide details of the conversion of laser energy to acoustic energy. Fig. 4 shows the evolution of the spectrum of the acoustic pulse emitted by the laser filament as a function of the longitudinal filament position. The laser propagates towards +z direction. It is seen from Fig. 4 that the acoustic bandwidth at z=0 increases from 0.35 MHz (Fig. 4a) to 1.1 MHz (Fig. 4b) as the focal lengths increase from 5 cm to 5 m at a given laser energy. This phenomenon might be explained by the fact that such a strong focusing ($f$ =5 cm) promotes the generation of a very dense and short plasma at the leading edge of the laser pulse, which exerts a strong defocusing effect on the peak and tail of the laser pulse, leading to the decrease of the optical intensity and a less efficient energy deposition. For further increases in focal length, it shows a different scenario. As the change of focal length from 5 m to 10 m and 20 m, the acoustic bandwidth gradually decreases to 0.7 MHz (Fig. 4c) and 0.5 MHz (Fig. 4d). It illustrates that further increasing focal length leads to an overall decrease of energy deposition, which is consistent with the existing research results [32].

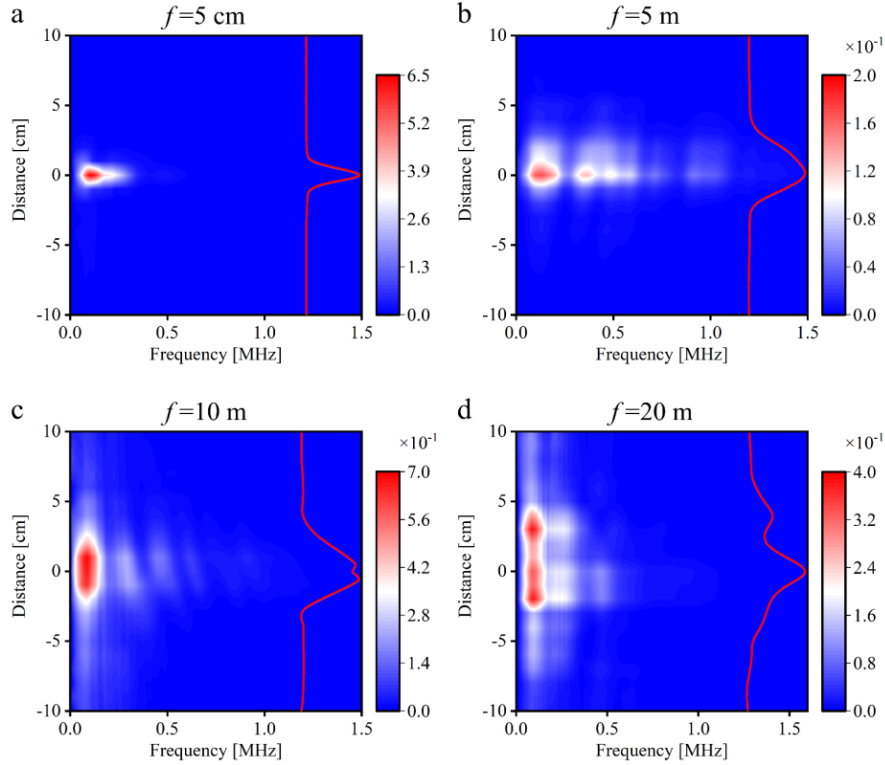

Fig. 4. Evolution of the acoustic spectrum along the femtosecond laser filament for different focal lengths of 5 cm (a), 5 m (b), 10 m (c), and 20 m (d).

The distribution of the acoustic amplitude along the optical filament is obtained by integrating the spectrum in Fig. 4 and shown by the red line on the right side of each plot. Since the heating of the ambient material by the laser filament occurs much more quickly than the characteristic time of air hydrodynamic response, the heating process can be considered as an isochoric process. The deposited energy $\Delta U$ can be written as:

$$\Delta U = c_v n_0 \int_V (T(\vec{r}) - T_{air}) dV$$
$$= \frac{c_v}{k_B} \int_V (p(\vec{r}) - p_{air}) dV \quad (1)$$

where $c_v$ is the isochoric heat capacity of an air molecule and $n_0$ is the air density at $p_{air} = 1.013 \times 10^5$ Pa and $T_{air} = 300$ K. Assuming the initial pressure profile is uniform in the cross section of the filament and only varies along the laser propagation direction ($z$ direction), this equation can be simplified as:

$$\Delta U \approx \frac{\pi c_v}{k_B} \int_L (p_{max}(z) - p_{air}) r_0(z)^2 dz \qquad (2)$$

where $L$ is the length of the filament and $r_0(z)$ is the half width at half maximum (HWHM) of the temperature profile at $z$. Assuming that $r_0(z)$ slowly varies along the filament, Eq. (2) can be further express as:

$$\Delta U \propto \int_L (p_{max}(z) - p_{air}) dz \qquad (3)$$

According to Eq. (3), the acoustic amplitude is proportional to the local energy absorption (or energy deposited per unit length) by the laser pulse, and the $z$-evolution of the acoustic amplitude gives the profile of the deposited energy along $z$. We recorded the variation of the acoustic bandwidth along the filament at a focal length of 5 m. Results are plotted in Fig. 5 (red line), compared with the acoustic amplitude profile (blue line). The acoustic spectrum at $z=0$ shows a bandwidth of 1.1 MHz and gradually decreases to 0.1 MHz as it is gradually close to the ends. It is interesting to note that acoustic bandwidth and the acoustic amplitude behave quite similarly showing the relation between the acoustic spectrum and the deposited laser energy. Due to the dependence between the pulse duration and pulse spectrum, the broader the acoustic spectrum is, the shorter the acoustic pulse duration is. However, since in time domain the acoustic pulse waveform is always composed of a series of acoustic oscillations and hard to exactly determine the pulse duration, it is more convenient to judge the filament distribution and dynamic process by the acoustic bandwidth in frequency domain.

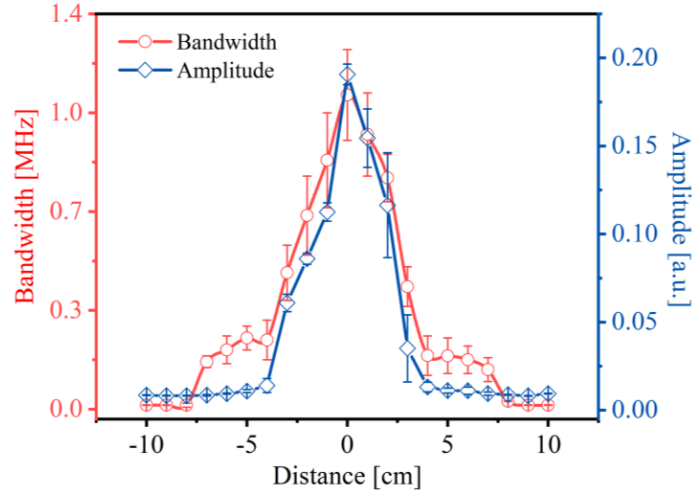

Fig. 5. Bandwidth of the acoustic pulse spectrum (red line) and the acoustic amplitude (blue line) along the optical filament. The focal length of the focusing lens used here is 5 m.

## 4. Conclusion

In summary, the broadband microphone was employed to quantitatively measure the acoustic emission of femtosecond laser filament for different focal lengths and pulse energies. Acoustic features related to the laser energy deposition were identified and discussed. The results showed that acoustic power presents a squared dependence on the laser energy $\overline{W} \propto I^2$, which is consistent with the gas absorption via non-resonant two-photon rotational Raman excitation in air molecules. The focal length dependence of the bandwidth of the acoustic pulse is related to the energy deposited under different focusing conditions. The spectrum of the acoustic pulse is also an indicator of the filament intensity distribution and showed a significant positive correlation between the acoustic bandwidth and the laser energy deposition. The studies have established a relationship between the acoustic parameters and the laser pulse parameters and focusing conditions. These findings are helpful for studying the plasma properties and dynamics processes through acoustic parameters and allow the optimization of remote applications.

**Funding.** National Key Research and Development Program of China (2018YFB0504400) and Fundamental Research Funds for the Central Universities (63223052).

**Disclosures.** The authors declare no conflicts of interest.

**Data availability.** Data underlying the results presented in this paper are not publicly available at this time but may be obtained from the authors upon reasonable request.